\documentstyle{mn}
\input{epsf}
\newcommand{\ba}{\begin{eqnarray}}
\newcommand{\ea}{\end{eqnarray}}
\newcommand{\be}{\begin{equation}}
\newcommand{\ee}{\end{equation}}
\newcommand{\nn}{\nonumber \\}
\newcommand{\vk}{\vec{k}}

\newcommand{\vr}{\vec{r}}
\newcommand{\vx}{\vec{x}}
\title{Large-scale bias in the Universe:  bispectrum method}
\author[S. Matarrese, L. Verde \& A.F. Heavens]
{S. Matarrese$^{1}$, L. Verde$^{1,2}$, 
A.F. Heavens$^{2}$\\
$^{1}$ Dipartimento di Fisica {\em Galileo Galilei}, Universit\`{a} di
Padova, via Marzolo 8, I-35131 Padova, Italy\\
$^{2}$ Institute for Astronomy, University of Edinburgh, Royal Observatory, 
Blackford Hill, Edinburgh EH9 3HJ , United Kingdom\\}
\begin{document}
\maketitle
\begin{abstract}
Evidence that the Universe may be close to the critical density, 
required for its expansion eventually to be halted, comes principally from
dynamical studies of large-scale structure.   These studies 
either use the observed peculiar velocity field of galaxies directly, or 
indirectly
by quantifying its anisotropic effect on galaxy clustering
in redshift surveys.   A potential difficulty with both such approaches 
is that the density parameter $\Omega_0$ is obtained only 
in the combination $\beta = \Omega_0^{0.6}/b$, if linear 
perturbation theory is used.  The determination of 
the density parameter $\Omega_0$ is therefore compromised by the lack of 
a good measurement of the bias parameter $b$, which relates the clustering of
sample galaxies to the clustering of mass.

In this paper, we develop an idea of Fry (1994), using second-order
perturbation theory to investigate how to measure the bias parameter on 
large scales.   The use of higher-order 
statistics allows the degeneracy between $b$ and $\Omega_0$ to be lifted,
and an unambiguous determination of $\Omega_0$ then becomes possible.
We apply a likelihood approach to the bispectrum,  
the three-point function in Fourier space.    
This paper is the first step in turning the idea into a practical 
proposition for redshift surveys, and is principally concerned with 
noise properties of the bispectrum, which are non-trivial.  
The calculation of the required bispectrum covariances involves 
the six-point function,  including many noise terms, for which we have 
developed a generating functional approach which will be of value 
in calculating high-order statistics in general.
\end{abstract}
 

\section{Introduction}

The assumption that bright galaxies are {\em biased} tracers of
the mass distribution has featured strongly in theories
of galaxy and structure formation in recent years.  The idea that galaxies 
are biased even 
on large scale has been studied for example by Peacock \& Heavens (1985) and
Bardeen et al. (1986), where galaxies are hypothesized as forming at 
high peaks of the density field filtered on relatively small scales.
The concept of bias was then further extended to the extreme case where 
galaxies are painted on arbitrarily on the mass distribution.
However, some constraints can be imposed,  when the galaxy density is an 
arbitrary local function of the mass density.
In this case, the bias (here defined in terms of the two-point correlation 
function) must be a monotonic function of the spatial separation (Coles 1993)
if it derives from a Gaussian field.
Numerical experiments show that this also holds for the power spectrum
(Mann, Peacock \& Heavens 1997).  

It has been suggested that the bias mechanism can be modulated by
environment dependent effects, as for example in the Cooperative
Galaxy Formation model (Bower et al. 1993).
The net result of this modification of the standard scheme is that the 
relationship between the density fluctuation field $\delta(\vx)$ and the 
galaxy fluctuation field $\delta _g (\vx) $ becomes non-local and 
the bias turns out to be scale-dependent.  Other bias possibilities include
dynamical friction effects, allowing galaxies to settle in clusters 
(Couchman \& Carlberg 1992).   Note that even if the physical mechanism 
responsible for the bias operates only on relatively small scales, 
(e.g. within cluster cores), the bias parameter may differ from unity
on much larger scales (Mann, Peacock \& Heavens 1997), with long wavelength
modes having to have enhanced amplitude to fit the higher peaks of density 
in clusters.
If the bias arises physically in the epoch of galaxy formation, 
it will evolve in time, and will approach unity if galaxy numbers are
preserved (Fry 1996, Nusser \& Davis 1994).  This condition may be broken by
mergers (e.g. Matarrese et al. 1997), but also by luminosity evolution, if the 
sample is defined by a flux criterion, so it is safest to make no assumptions 
about bias evolution.

There are clearly many ideas, but we have really very little idea of how
bias works and evolves in practice, and there is therefore strong
motivation to find ways to measure it empirically from the galaxy 
distribution.    The main motivation, however, comes from the desire
to split the degeneracy between $\Omega_0$ and $b$ which arises in
linear perturbation theory from dynamical studies of large-scale structure.
Both direct studies of the peculiar velocity field (e.g. Dekel et al. 1993)
and studies of the distortion which peculiar velocities give to redshift-space
galaxy maps (Kaiser 1987, Hamilton 1992, Fisher, Scharf \& Lahav 1994, 
Heavens \& Taylor 1995, Ballinger, Heavens \& Taylor 1995) yield a value of
$$
\beta \equiv\frac{\Omega_0^{0.6}}{b},
$$
but not $\Omega_0$ or $b$ separately.
The ignorance of the bias parameter compromises all conclusions about 
the real prize -- $\Omega_0$, from large-scale structure studies.

With good data, probably available in the next few years from the new 
generation of galaxy surveys, we show that the bias parameter can be 
measured by studying higher-order characteristics of the density field.
More specifically this is done by analyzing the 
bispectrum: the Fourier transform of the three-point correlation 
function (e.g. Frieman \& Gazta\~naga 1994).  
This statistic has the advantage that the determinations
of the bispectrum can be made uncorrelated on different scales, or
if not, their correlation properties can be calculated, and the 
choice of modes to analyse can be made to maximize signal-to-noise.
These allow the bias parameter to be estimated via likelihood methods,
importantly allowing error bars to be placed on the parameter estimates.
The higher-order statistics work by exploiting the fact that gravitational
instability skews the density field to high densities as it evolves.
This behaviour can also be mimicked by non-linear biasing,  but these 
two effects can be separated by the use of shape information: 
essentially non-linear biasing of a truly Gaussian field will lead to 
different shaped structures from a non-Gaussian field arising from
gravitational instability. 

It is worth making a few remarks about the advantages of performing
this sort of analysis in Fourier space rather than via high-order
correlation functions in real space.  The main advantage, which is 
shared by the power spectrum, is that the estimates of the correlation
functions in Fourier space can be made uncorrelated (or at least their
correlation properties can be readily calculated), with enormous benefits
for estimation of parameters and error bars.  The second advantage is that
the real-space determinations of high-order correlations are really only
known well in the non-linear regime, where perturbation theory may
not apply.  In Fourier space, there is a clear 
separation of scales where perturbation theory works and breaks down.

Translating this idea into a practical algorithm requires a 
number of issues to be tackled, including the effects of a varying
selection function and redshift distortions, and the non-trivial
issue of what shot noise terms appear in the required six-point function.  
This paper is principally concerned with the error analysis;
a subsequent paper will present a study of the other issues.

The plan of this paper is as follows. In Section 2 we introduce the 
second-order perturbation expansion for the (unknown) matter density 
field and the (observable) galaxy density field.
The necessary mathematical techniques to calculate the $N$-point functions, 
including shot noise,  are given 
in Section 3.  Sections 4 discusses optimal methods and practical 
implementation to a numerical simulation of the galaxy distribution, 
and Section 5 discusses the results and the possibilities of practical 
application to real data.

\section{Second order approximation}

As Fry (1994) pointed out, since the degeneracy between $\Omega_0$ 
and $b$ is an intrinsic feature of the linear theory, one needs to 
go to second order in order to separate the parameters.

Under the assumption that the initial fluctuation are Gaussian and that 
structure grows by gravitational instability, the 
three-point correlation function is intrinsically a second-order quantity, 
and should be detectable in the mildly non-linear regime where 
second-order perturbation theory is a reliable description. 

The data we will use to constrain the bias parameter will be the real
part of the three-point function in $k$-space:
\begin{equation}
D_\alpha = {\rm Re}(\delta_{\vk 1}\delta_{\vk 2}\delta_{\vk 3})  ,
\end{equation}
for closed triangles. $\alpha$ is shorthand for the triangle
specified by some triplet of k-vectors.  $\delta_{\vk} \equiv 
\int d^3\vec{r}\,\delta(\vec{r})
\exp(-i\vec{k} \cdot \vec{r})$
is the Fourier transform of the fractional overdensity field $\delta(\vec{x})
\equiv n(\vec{x})/\bar{n}-1$, where $n$ is the number density of galaxies,
and $\bar n$ its mean.
$D_\alpha$ is related to the bispectrum $B(\vec{k_1},\vec{k_2},\vec{k_3})$
by
\be
\langle \delta_{\vk 1}\delta_{\vk_2}\delta_{\vk_3} \rangle = 
(2\pi)^3 B(\vk_1,\vk_2,\vk_3) \delta^D
(\vk_1+\vk_2+\vk_3) = \langle D_\alpha \rangle , 
\ee
where $\delta^D(\vec{k})$ is the three-dimensional Dirac delta function 
and the $\langle \rangle$ indicates ensemble average, or, by the 
ergodic theorem, the average over a large volume.
In what follows, we may occasionally refer loosely to $D_\alpha$ as the 
bispectrum.

We expand the density field to second order as:
\begin{equation}
\delta(\vec{x})= \delta ^{(1)} (\vec{x}) + \delta ^{(2)} (\vec{x}) ,
\end{equation}
where $\delta ^{(2)}$ is $O(\delta ^{(1)2})$ and represents departures from
Gaussian behaviour. 
     
Note that what we are dealing here is not a mildly nonlinear field, but 
a highly nonlinear field which is filtered on some smoothing scale.
The operations of smoothing and gravitational evolution do not commute,
so there is some possibility that perturbation theory might be inaccurate.
However, this problem is not unique to second-order perturbation theory: it
exists in even for the
evolution of the power spectrum, which for practical cases can be treated 
well by perturbation theory.   Experiments on the bispectrum by others
and ourselves (see section 4) show that perturbation theory for the
implicitly filtered fields works well if the smoothed field is not too 
nonlinear.

\subsection{Unbiased case}

For an unbiased distribution (or for the matter distribution) we can 
expand \mbox{$\langle\delta_1 \delta_2 \delta_3\rangle$} 
to second order, obtaining:
\ba
\langle\delta_1 \delta_2 \delta_3\rangle & \simeq & \langle\delta_1^{(1)} 
\delta_2^{(1)} \delta_3^{(1)}\rangle\nn
& & + \langle\delta_1^{(1)}\delta_2^{(1)}\delta^{(2)}_3\rangle
+ cyclic\ terms (231,312) . \nn
\ea
Only the second term of this survives.
Since we will work in Fourier space we quote directly the expression for 
$\delta^{(2)}_{\vec{k}}$, the Fourier transform of $\delta^{(2)}(\vec{x})$
(Catelan et al. 1995):
\begin{equation}
\delta^{(2)}_{\vec{k}}= {1\over (2\pi)^3} \int  d^3 k_a d^3 k_b 
\delta^D(\vec{k}_a+ \vec{k}_b - 
\vec{k}) J(\vec{k}_a , \vec{k}_b) \delta^{(1)}_{\vec{k}_a} 
\delta^{(1)}_{\vec{k}_b} ,
\end{equation}
where
\ba
J(\vec{k_1},\vec{k_2},\Omega_0) & = & 
1-B(\Omega_0)+\nn
& &  \!\!\!\!\!\! \frac{ \vec{k_1}\cdot \vec{k_2} } 
{ 2k_1k_2 } \left( \frac{k_1}{k_2} + \frac{k_2}{k_1} \right) 
+B(\Omega_0) \left( \frac{ \vec{k_1}\cdot \vec{k_2}}{k_1k_2} \right) ^2.\nn
\ea
$B$ is a function which is $2/7$ for an Einstein-de Sitter universe, in which 
case the expression above reduces to  the form originally obtained by Goroff 
et al. (1986). 
The useful feature of this is that $J$ is almost independent of
$\Omega_0$, as seen in Fig. 1, so we can use the Einstein-de Sitter 
value with minimal error.

\begin{figure}
\begin{center}
\setlength{\unitlength}{1mm}
\begin{picture}(90,70)
\includegraphics{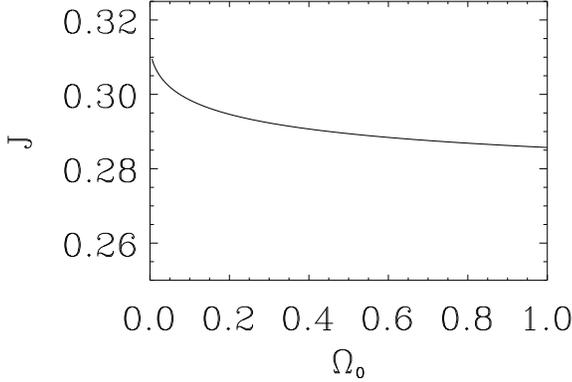}
\end{picture}
\end{center}
\caption{The weak dependence of $J$ on $\Omega_0$ for equilateral triangles.}
\end{figure}

The bispectrum for the unbiased case, in the absence of shot noise,
is therefore
\ba
\langle \delta_{\vk 1}\delta_{\vk 2}\delta_{\vk 3} \rangle & = &
(2\pi)^3\left[2J(\vk_1,\vk_2)P(k_1)P(k_2)+ cyc. (23,\ 13)\right]\times \nn
& & \delta^D(\vk_1+\vk_2+\vk_3) ,
\ea
where the linear power spectrum is defined in terms of the orthogonality of
the transform coefficients, arising from homogeneity:
\be
\langle \delta_{\vk 1} \delta_{\vk 2} \rangle = (2\pi)^3 P(k) \delta^D 
(\vk_1+\vk_2).
\ee
For practical cases, the transform is made in a finite box, when the
Dirac delta function is modified (see Section 3.3).

\subsection{Biased case}

We assume that the biased (galaxy) density is a local function of the 
unbiased (matter) density. Galaxy properties will be identified by
the subscript $g$;  matter fields have no subscript.  
We make a Taylor expansion of the galaxy density to second order 
in $\delta^{(1)}$, so for consistency we must include an extra quadratic bias
term $b_2$:
\ba
\delta_g(\vec{x})& = &f[\delta (\vec{x})]\nn
 & \simeq & b_1 \delta(\vec{x})
+\frac{1}{2} b_2 
\delta^2(\vec{x}) \nn
& \simeq & b_1 \delta^{(1)}(\vec{x}) + b_1 
\delta^{(2)}(\vec{x})+ \frac{1}{2} b_2 \delta^{(1)2} (\vec{x}).\nn
\ea
The linear bias parameter $b_1$ is to be identified with $b$.  Notice
that the expression above is not correctly normalised:  a term 
$b_2 \langle\delta^{(1)2}\rangle/2$ needs to be subtracted so that the biased 
field has zero mean.  As we will be working in Fourier space, this extra term
is irrelevant for all except the $\vk=\vec{0}$ mode, and will be 
ignored\footnote{It is not obvious that a physical mechanism will lead to a 
functional dependence which can be expanded in a Taylor series.  Examples
are the `weighted bias' model studied by Catelan et al.
(1994) and the `censoring bias' studied by Mann, Peacock \& Heavens (1997). 
It is an open question whether such schemes might be
approximated by a continuous $\delta_g(\delta)$ relation when smoothed on
larger scales.}.

Following the same procedure as in the unbiased case we obtain to second order
\begin{eqnarray}
\langle\delta_{g1} \delta_{g 2} \delta_{g3}\rangle & = & b_1^3 \langle 
\delta^{(1)}_1 \delta^{(1)}_2 \delta^{(1)}_3\rangle\nn
  & + &  b_1^3 \langle\delta^{(1)}_1 \delta^{(1)}_2 \delta^{(2)}_3\rangle + 
cyc.\nn
   & + & \frac{b_1^2 b_2}{2} \langle \delta^{(1)}_1 \delta^{(1)}_2 
\delta^{(1)2}_3 \rangle + cyc.\nn
\end{eqnarray} 
Once again, the first term vanishes.
We can re-write the last term by exploiting the cumulant 
expansion theorem for the four-point function 
obtaining the non-vanishing terms
\be
\langle\delta^{(1)}_1 \delta^{(1)}_2\rangle\langle\delta_3^{(1)2}\rangle+
2\langle\delta^{(1)}_1 
\delta^{(1)}_3\rangle\langle\delta^{(1)}_2 \delta^{(1)}_3\rangle.
\ee
Therefore to second order the galaxy bispectrum is:
\ba
& &\langle\delta_{g \vec{k}_1} \delta_{g \vec{k}_2} 
\delta_{g \vec{k}_3}\rangle =\nn
& & (2\pi)^3
\left\{ b_1^3 \left[P(\vec{k}_1) P(\vec{k}_2) 2 J(\vec{k}_1,\vec{k}_2)+ 
cyc.\right] \right. \nn
& & +\left. b_1^2 b_2 \left[ P(\vec{k}_1) P(\vec{k}_2) + {cyc.} \right] 
\right\}
\delta^D (\vec{k}_1 +\vec{k}_2 +\vec{k}_3).\nn
\label{bisp}
\ea
From this we see how the bispectrum may be used to estimate the bias 
parameters.  $J$ is almost independent of $\Omega_0$, so a set of 
triangles in k-space can in principle allow us to measure the bias 
parameters $b_1$ and $b_2$.  It is important to realise that
it is necessary to consider triangles of different shapes to lift a
new degeneracy between $b_1$ and $b_2$, since triangles
of a common shape have the same $J$ and (\ref{bisp}) then gives
information only on the combination $b_1^2(2 b_1 J+b_2)$.  The same sort
of degeneracy has been noted for the 3-point function in real space 
(see Frieman \& Gazta\~naga 1994).  The strong dependence on shape has
been illustrated (e.g. Jing, B\" orner \& Valdarnini 1995), but there 
appears to be little work on using different shapes to lift the degeneracy.   
We return to the physical reasons for having to take more than one shape
in the discussion.

We wish to estimate $b_1$ and $b_2$ by maximum 
likelihood,
so we need an estimate of the shot noise, and the covariance between 
bispectrum estimates.   We turn to this in the next section, but here
note that the bispectrum (strictly defined in terms of the mean) is a
real quantity.  This is a result of isotropy - the bispectrum does not depend
on which direction round the triangle is taken.  Accordingly we use only
the real parts of $\delta_{g \vec{k}_1} \delta_{g \vec{k}_2} 
\delta_{g \vec{k}_3}$ as our data, as the imaginary parts just introduce 
more noise. 

Since $P(\vec{k})$ is unknown, we write (\ref{bisp})  
in terms of potentially observable quantities, the linear galaxy power 
spectrum $P_g = b_1^2 P$ (equal to the observed power spectrum at 
adequate order for second-order perturbation theory).:
\ba
& &\langle\delta_{g \vec{k}_1} \delta_{g \vec{k}_2} 
\delta_{g \vec{k}_3}\rangle =\nn
& & (2\pi)^3
\left\{c_1 \left[2 P_g(\vec{k}_1) P_g(\vec{k}_2) J(\vec{k}_1,\vec{k}_2)+ 
cyc.\right] \right. \nn
& & + c_2 \left. \left[ P_g(\vec{k}_1) P_g(\vec{k}_2) + {cyc.} 
\right] \right\} \delta^D (\vec{k}_1 +\vec{k}_2 +\vec{k}_3),\nn
\label{bisp2}
\ea
where
\be
c_1 \equiv {1\over b_1};\qquad c_2 \equiv {b_2\over b_1^2} .
\ee
Under the assumption of a uniform prior for the quantities $c_1$ \& $c_2$,
the {\it a posteriori\/} probability for the parameters is the likelihood:
\ba
{\cal L}(c_1,c_2)& = &\frac{1}{(2 \pi )^{\frac{M}{2}} 
({\rm det} C)^{\frac{1}{2}}}\nn 
& & \exp{\left[-\frac{1}{2} \sum_{\alpha \beta}(D_{\alpha } - \mu _{\alpha}) 
C_{\alpha \beta}^{-1} ( D_{\beta} - \mu_{\beta }) \right]} ,\nn 
\label{like}
\ea
where the means of the $M$ data are $\mu_\alpha(b_1, b_2) 
\equiv \langle D_\alpha \rangle$ and  
$C_{\alpha \beta} \equiv \langle(D_{\alpha }- \mu_{\alpha })(D_{\beta} - 
\mu _{\beta })\rangle$
is the covariance matrix, which is also a function of the bias parameters.
The use of a Gaussian likelihood function is justified if there are many
contributing data, whether or not they themselves have Gaussian distributions,
provided the sensitivity to the parameters is through the mean rather than
the variance.   It is straightforward to show that, by using the mean of
many observed data drawn from the same distribution, the likelihood is
identical (by the Central Limit Theorem) as the Gaussian likelihood for 
the individual data, with the same variance and mean.   In our case
the variance is only weakly dependent on the parameters, through the shot
noise, so our use of a Gaussian likelihood function is quite justified.

Note that our data are real $D_\alpha = Re(Z_\alpha) \equiv
Re(\delta_{\vec{k}1} \delta_{\vec{k}2} \delta_{\vec{k}3})$, and we can
write the covariance matrix for $D_\alpha$ in terms of the complex $Z_\alpha$:
\be
\langle D_{\alpha }D_{\beta }\rangle= \frac{1}{2} Re\left[ 
\langle Z_{\alpha }Z_{\beta }\rangle+
\langle Z_{\alpha }Z^*_{\beta }\rangle \right].
\label{ZZstar}
\ee
It should be obvious that we require the six-point function to
calculate the covariance matrix, and we also need to include shot noise
in the bispectrum estimates.  The former is sufficiently complicated
by the presence of a very large number of noise terms that we present
a general method for doing calculations of this nature in the next section.

\section{Statistics of random fields and derivation of the Pentaspectrum}

This section presents a mathematical method for calculating
$N$-point distributions in Fourier space for continuous and discrete fields,
and deal also with the realistic case of a galaxy survey with a non-uniform
number density (arising, for example, from a flux-limited catalogue).  These
distributions are required to calculate the covariance properties of the 
bispectrum.   A reader interested principally in the application of the
bispectrum may omit this at a first sitting.

The complete description of a random field is given by the {\it 
probability distribution functional\/} ${\cal P}$, which should be thought as 
the continuous limit of the joint probability distribution in N points, as 
$N \rightarrow \infty$. To describe the 
statistics of a general random field $f(\vec{x})$, 
we need to know the functional ${\cal P}[f(\vec{x} )]$, which 
specifies the distribution of the possible values of $f$ at each point. 
By statistical homogeneity ${\cal P}[f(\vec{x} )]$ will be independent of 
$\vx$. 
Consider then the {\it partition functional\/} or {\it generating functional 
of the correlation functions\/} ${\cal Z}$ defined as a functional 
integral, as follows: 
\ba
{\cal Z} [{\cal J}(x)] & \equiv & \int {\cal D} [f] {\cal P} 
[f ] \exp \left(i \int  d^3x {\cal J}(\vec{x})f(\vec{x})\right)\nn
& = &
\left\langle \exp \left[ i 
\int  d^3 x {\cal J}(\vec{x}) f(\vec{x}) \right] \right\rangle ,
\ea
where ${\cal J}(\vec{x})$ is an external source perturbing the underlying 
statistics and ${\cal D} [f]$ is a suitable measure such that the 
probability distribution is correctly normalized to 1: $\int {\cal D}
[f(\vec{x})] {\cal P} [f (\vec{x})] =1$.

From this function we can recover the {\it correlation functions\/} $\mu_N$ 
of the distribution and the {\it connected correlation functions\/} or 
$\kappa_N$ as follows: 

\begin{itemize}

\item From the {\it generating functional of the correlation functions}
\begin{equation}
\mu_N \equiv \langle f(\vec{x}_1)\ldots f(\vec{x}_N)\rangle 
=i^{-N} \frac{ 
\delta^N {\cal Z}[{\cal J}]}{ \delta {\cal J}(\vec{x}_1)\ldots
\delta {\cal J}(\vec{x}_N)} ,
\label{gf}
\end{equation}
evaluated at ${\cal J}=0$.\\

\item From the {\it generating functional of the connected (or `reduced') 
correlation functions}
\be
{\cal K}[{\cal J}] \equiv \ln {\cal Z}[{\cal J}] ,
\ee

\ba
\kappa_N(\vec{x_1}\ldots\vec{x_N}) & = & \langle f(\vec{x}_1)\ldots
f(\vec{x}_N)\rangle_{\rm connected}\nn
& = & i^{-N} \frac{\delta^N {\cal K}[{\cal J}]}
{\delta {\cal J}_1\ldots\delta {\cal J}_N} ,\nn 
\ea
evaluated at ${\cal J}=0$. 

\end{itemize}
From the above definitions it follows that a complete 
characterization of the
statistics, i.e. the complete knowledge of the partition functional,
requires the knowledge of the correlation functions of all
orders.  In fact it is usually possible to recover the partition functional
as
\ba
{\cal Z}[{\cal J}] & = & 1+ \sum_{n=1}^{\infty} \frac{i^n}{n!} 
\int d \vec{x_1}\ldots d \vec{x_n} 
\mu_n(\vec{x}_1,\ldots,\vec{x}_n) \times \nn
& & {\cal J}(\vec{x_1}) \cdot \cdot 
\cdot {\cal J}(\vec{x_n})
\ea
and similarly for ${\cal K}[{\cal J}]$ in terms of the connected 
correlation functions, 
\ba
{\cal K}[{\cal J}] & = & \sum_{n=1}^{\infty} \frac{i^n}{n!} 
\int d \vec{x_1}\ldots d \vec{x_n} 
\kappa_n(\vec{x}_1,\ldots,\vec{x}_n) \times \nn
& & {\cal J}(\vec{x_1}) \cdot \cdot 
\cdot {\cal J}(\vec{x_n}) .
\ea

The expansion of correlation functions of order
N in terms of connected correlation functions of order M$\leq$N, i.e. the 
Cumulant (or Cluster) Expansion (e.g. Kendall \& Stewart 1977, van 
Kampen 1992), follows trivially from equations (21) and (22). 
The cluster expansion is widely used in statistical mechanics where it 
constitutes the basis for approximation schemes in studying slightly non-ideal 
many-body systems like gases or fluids (Rice \& Gray 1965). 
A review of the use of functional integration in cosmology can be found 
e.g. in Fry (1984), Bertschinger (1992), Matsubara (1995). 
It is worthwhile, at this point, to make some remarks on applicability.
In dealing with slightly imperfect fluids it is assumed that the 
interaction falls off faster than $\frac{1}{r^3}$, this allows one to 
neglect the connected higher-order terms; in the case of gravity 
we must be aware that this is not generally true.
The connected correlation functions $\kappa_N$, which are closely related to
the connected Green's functions of quantum field theory (e.g. Ramond 1989),
are defined by the above procedure in such a way that any
separable contribution of the type $\langle f^M\rangle \langle
f^{N-M}\rangle$ has been removed. As a consequence, if any subset of the
N points is set to infinite separation the connected correlation function
of order N goes to zero. These connected correlation functions of order N 
for the density fluctuation field are often simply called 
`correlation functions' in the cosmological literature (e.g. Peebles 1980). 
An important property of these functions is that if the underlying statistics 
is Gaussian all the connected correlations of order higher than 2 vanish. 
Finally, when the covariance functions (and the connected analogous) are 
evaluated for $\vec{x}_1 = \vec{x}_2 = \ldots =\vec{x}_N $, respectively 
the {\it moments\/} and the {\it cumulants\/} are obtained.

Continuous random fields and point processes (random distributions of
discrete points) are both cases of random processes. It is possible to
relate discrete and continuous distributions as we will see in the
next paragraphs.  

\subsection{Discrete and continuous distributions}
As already mentioned, the theory predicts the statistical properties
of the continuous matter distribution, while observations are concerned
with the galaxy distribution, which is discrete. 
Different biasing schemes have been 
introduced to relate the (theoretical) distribution of relative 
mass density 
fluctuations to the (observed) distribution of galaxies, but the true 
relation is still unknown. However, independently of the presence of 
bias, it is possible to relate formally one to the other.

In order to recover a continuous distribution from a discrete one, define 
for the discrete process the analogue of the density field, the number 
density field, as a sum over Dirac delta functions:
\ba
f(\vx)=n(\vec{x}) & = & \bar{n} [1+ \delta (\vec{x})] \equiv
\sum_i \delta^D(\vec{x}-\vec{x}_i),\nn  
\bar{n} & = & 
\left\langle \sum_i \delta^D(\vec{x}-\vec{x}_i)\right\rangle ,
\ea
where $\bar{n}$ is the mean number density.

Conversely, a simple way to generate a discrete distribution from a 
continuous one is to assume 
Poisson sampling (Peebles 1980, Bertschinger 1992, Fry 1985) by 
placing, at each volume element $\delta V$ in the sample 
region, a point with probability $\delta {\it P}=\rho (\vec{x}) \delta V$,
where $\rho(\vx)$ is the local mean density.
The outcome is a double stochastic process, with one level of randomness 
coming from the random field and the second from  the Poisson 
sampling\footnote{This procedure to account for discreteness is 
a model,  and may not hold in practice. An example arises in Voronoi
tesselations (see e.g. Williams, Peacock \& Heavens 1991), where objects are 
placed at the intersections of Voronoi
planes.  This case can even result in sub-shot noise power.}.
It is then possible to  take into account the effect of discreteness by 
performing a simple substitution in the partition 
functional or in the generating functional for the continuous case. 

In the Poisson model if the probability of finding one object in an 
infinitesimal volume element $\delta V$ at position 
$\vec{x}$ is $P_1=\rho(\vec{x}) 
\delta V$, then the probability of finding no objects is 
$P_{0} =1-\rho (\vec{x}) \delta V$ and the probability of finding more than 
one object is an infinitesimal of higher order. To obtain the probability 
generating {\em functional} for the discrete process, we need to generalize 
this procedure to infinitely many small volumes. With this aim, 
we start from a given realization of the stochastic process and 
divide the volume in an infinite set of cells with individual volumes 
$\delta V_\ell$. Then we consider the joint 
probability $P_{\{ N_\ell \}}$ of finding $N_1$ objects in the volume $\delta V_1$, 
$N_2$ in the volume $\delta V_2$, etc.. and we 
define the generator of discrete counts in that realization as
\ba
{\cal Z}[{\cal J}] & = & \sum_{\{ N_{\ell} \}} P_{\{ N_{\ell} \}} 
e^{i\sum_{\ell} N_{\ell}{\cal J}_{\ell}}\nn
& = & \prod_{\ell} \left[1-\rho(\vec{x}_{\ell}) \delta V_{\ell} + 
\rho(\vec{x}_{\ell}) 
\delta V_{\ell} e^{i{\cal J}(\vec{x}_{\ell})} \right] \nn
& \cong & \prod_{\ell} \exp \left[(e^{i{\cal J}(\vec{x}_{\ell})} - 1)\rho 
(\vec{x}_{\ell}) \delta V_{\ell}\right] .
\ea
We then let $\delta V_\ell \to 0$ in the given realization and average over  
the statistical ensemble of the underlying continuous field. This gives the 
required generating functional for the discretized process 
\be
{\cal Z}^d[{\cal J}] = \left\langle \exp  \int d^3x \left[e^{i{\cal J}(\vec{x})}
-1 \right]
\rho (\vec{x}) \right\rangle.
\ee
Accounting for the effect of graininess therefore amounts to the substitution 
\begin{equation}
i{\cal J} \rightarrow \left( e^{i{\cal J}} -1 \right) 
\end{equation}
in the generating functional for the continuous process. This result is 
a generalization of the standard procedure leading to the generating 
{\em function} for the moments of discrete counts in cells (e.g. Fry 1985). 
The practical utility of this method is that it permits calculation of 
correlated counts in different cells, according to the procedures described 
in Section 3.2 and Section 3.3 below. 

Actually, there is a subtlety to point out here:
the discrete process that we can observe is the one connected with the 
underlying continuous density distribution $\rho (\vec{x})$, but we want 
to study the statistics of the fractional over-density field 
$\delta=\rho(\vec{x})/\bar{\rho }-1$ which 
has several advantages, mainly the fact that it has zero mean.
To calculate its properties we have to modify slightly the above procedure.
We first notice that the discrete version of the random field $1+\delta$ 
is obtained by the substitution 
\be
i{\cal J} \rightarrow 
\left( e^{i{\cal J}/N}-1\right)N,
\label{discr2}
\ee
where $N$ is the average number of objects within the sampling volume, 
in the generating functional of $1+\delta$. 
Next, to subtract the terms coming from the mean value $1$, we use the fact
that the generating functional for the sum of two independent random fields 
is just the product of the individual generating functionals. The discrete 
generating functional for $\delta$ is therefore
\be
{\cal Z}[{\cal J}]_{\delta}= \exp \left\{- i \int d^3x {\cal J_{\rm cont}}(\vx)\right\} 
{\cal Z}[{\cal J}]_{1+\delta} 
\label{Z.delta} 
\ee
because the generating functional for a uniform and continuous 
field of value $1$ is simply $\exp\left[ i \int d^3x {\cal J_{\rm cont}}
(\vx)\right]$.

In the next section, we give explicit expressions for the source function 
${\cal J}$, in real and Fourier space, and for continuous and discrete 
distributions.

\subsection{$N$-point correlation functions in real and Fourier space}

The quantity that we are dealing with is a complex number $Z_{\alpha}$ 
given by 
$\delta_{\vec{k}_1} \delta_{\vec{k}_2} \delta_{\vec{k}_3}$ where the vectors
form a triangle: $\vec{k}_1+\vec{k}_2+\vec{k}_3=\vec{0}$.

The average of $Z_{\alpha}$ over the ensemble (of closed triangles) is 
the bispectrum, but the actual quantity $Z_{\alpha}$ fluctuates across 
the ensemble.
In order to quantify this fluctuation it would be valuable to estimate 
its covariance and this involves the quantity $\langle Z_{\alpha}
Z_{\beta}\rangle$ (see equation \ref{ZZstar}).

This quantity is nothing but the six-point correlation function (not 
just the irreducible or connected one)  in 
Fourier space (shot noise contribution included) in the particular case 
when the six vectors form two triangles:
\be
\langle\delta_{\vec{k}1} \delta_{\vec{k}2}.....\delta_{\vec{k}6}\rangle ,
\ee
where
$\vec{k}_1 +\vec{k}_2 +\vec{k}_3=\vec{0}$ and $\vec{k}_4 +\vec{k}_5+\vec{k}_6
=\vec{0}$.

The formalism illustrated in section 3 has been introduced for this reason.
Since we want to take into account the shot noise contribution, we need to work 
with the generating functional for the covariance functions of $1+\delta(x)$ 
which is:
\ba
{\cal Z}[{\cal J}] & = & \int{\cal D}(\delta){\cal P}(\delta) 
\exp \left(i \int d^3 x {\cal J}(\vec{x})\,[1+\delta(\vec{x})] \right)\nn
& = &
\left\langle{\exp 
\left\{ i \int d^3 x {\cal J}(\vec{x})\,[1+\delta(\vec{x})]  \right\}  
}\right\rangle . \nn
\ea
Instead of dealing with functional derivation, we may write the 
`external source' ${\cal J}$ in a form which allows the functional derivatives
to be replaced by normal differentiation (Matarrese, Lucchin \& Bonometto 
1986):
\begin{equation}
{\cal J}(\vec{x}) \equiv \sum^N_{m=1} s_m \delta^{D} (\vec{x} - 
\vec{x}_m).
\label{J}
\end{equation}
The covariance functions up to the $N^{th}$ order can then be obtained by the 
following differentiation:
\begin{equation}
\langle [1+\delta (\vec{x}_1)].....[1+ \delta(\vec{x}_N)]\rangle=i^{-N} 
\left[ 
\frac{ \partial^N {\cal Z} [{\cal J}]}{ \partial s_1....\partial s_N} 
\right] _{s_m=0} 
\end{equation}
and the corresponding connected parts are obtained performing the same 
kind of differentiation on $\ln {\cal Z}[{\cal J}]$.

Smoothing can be incorporated easily.  If the smoothed field is defined by
the convolution:
\be 
\delta_{\rm smoothed}(\vec{x}) \equiv \int d^3\vec{x}' \delta(\vec{x}')
W(\vec{x}-\vec{x}') ,
\ee
the statistics of the smoothed field may be obtained by the substitution
\begin{equation}
\delta ^D(\vec{x}-\vec{x}_m) \longrightarrow W (\vec{x}-\vec{x}_m)
\end{equation}
in equation (\ref{J}).
For a discrete process the generating 
functional ${\cal Z}^{d}[{\cal J}]$ 
is obtained with the substitution (\ref{discr2}).

In Fourier space, the $N$-point functions may be obtained by a straightforward 
change to the source function, which becomes the Fourier transform of the
real-space ${\cal J}$.  If $\widetilde{W}(\vec{k})$ is the transform of the 
smoothing function, then, for a continuous field,
\begin{equation}
{\cal J}_{\vec{k}}(\vec{x})=\sum_{m=1}^N s_m e^{-i \vec{k} \cdot \vec{x}}
\widetilde{W}(\vec{k}_m)
\label{sourcec}
\end{equation}
and in the discrete case:
\begin{equation}
{\cal J}_{\vec{k}}^d(\vec{x})=-iN \left\{ \exp \left[ \frac{i}{N}
\sum_{m=1}^N s_m e^{-i\vec{k}_m \cdot \vec{x}} \widetilde{W}(\vec{k}_m)
\right] -1 \right\}.
\label{sourced}
\end{equation}
In fact notice that the unsmoothed ${\cal J}_{\vec{k}}^d$ 
can be obtained from the 
${\cal J}^d$ 
just by substituting, in the exponential, the Dirac delta function by its 
Fourier transform. 

Our {\it ansatz\/ } for the generating functional is to assume that all 
correlations vanish above the 3-point function:
\begin{equation}
\begin{array}{c}
{\cal Z}[{\cal J}] =\\
\exp \left[ i \int d^3x {\cal J}(\vec{x})-\frac{1}{2}\int d^3x d^3 x^\prime 
{\cal J}(\vec{x}){\cal J}(\vec{x}^\prime ) \xi^{(2)}_{\rm conn.} 
(\vec{x},\vec{x}^\prime ) 
\right.\\
\left. - \frac{i}{6} \int d^3x d^3x^\prime d^3 x ^{\prime \prime} {\cal J}
(\vec{x}) 
{\cal J}(\vec{x}^\prime ) {\cal J}(\vec{x}^{\prime \prime} ) 
\xi ^{(3)}_{\rm conn.} (\vec{x},
\vec{x}^\prime ,\vec{x}^{\prime \prime} ) \right].
\end{array}
\end{equation} 

The validity of this form relies on the fact that we are assuming 
Gaussian initial fluctuations. In the linear regime all 
the connected correlation function of order 3 or higher vanish.
Allowing then for a quasi-linear evolution, and performing a 
second-order perturbation theory approximation, we allow $\xi^{(3)}$ 
to become non-zero, but we  still rely on all the higher-order 
irreducible correlation functions being negligible. 

Despite the conceptual simplicity of the algorithm introduced,  it is 
quite cumbersome to perform the
$6^{th}$ derivative of ${\cal Z}^d[{\cal J}]$, so for practical purposes
we use the cluster 
expansion and express the six-point correlation function in terms 
of the connected parts of lower or equal orders.
\ba
\langle\delta_1.....\delta_6\rangle^{d/c} & = & \nn
\langle\delta_1 \delta_2\rangle^{d/c}_{\rm conn.}\langle\delta_3 
\delta_4\rangle^{d/c}_{\rm conn.}\langle\delta_5 \delta_6\rangle^{d/c}_{\rm 
conn.} & + & \ldots 
\mbox{15 terms}\nn
  +\langle\delta_1 \delta_2\rangle^{d/c}_{\rm conn.}\langle\delta_3 \delta_4 
\delta_5 
\delta_6\rangle^{d/c}_{\rm conn.} & + & \ldots \mbox{15 terms}\nn
   +\langle\delta_1 \delta_2 \delta_3\rangle^{d/c}_{\rm conn.}\langle\delta_4 
\delta_5 
\delta_6\rangle^{d/c}_{\rm conn.} & + & \ldots \mbox{10 terms}\nn
   +\langle\delta_1\ldots \delta_6\rangle^{d/c}_{\rm conn.}.
\end{eqnarray}
The $d/c$ superscript stands for discrete or continuous case and it has 
been explicitly written to show that it is possible to treat the two 
cases, in configuration or Fourier space, on the same footing 
provided we make the corresponding substitution for the external source.

For our purposes the calculations have been made directly in Fourier 
space for the discrete case giving:
\be
\langle\delta_i \delta_j\rangle^d_{\rm conn.} =  (2\pi)^3\left[ P(k_i)+
\frac{1}{\bar{n}} \right] \delta ^D(\vec{k}_i + \vec{k}_j) ,
\ee
\ba
\langle\delta_l \delta_m \delta_n\rangle_{\rm conn.}^d &  = & (2\pi)^3  
\delta^D(\vec{k}_l+\vec{k}_m+\vec{k}_n)\times\nn
& & \!\!\!\!\!\!\!\!\!\!\!\!\!\!\!\!\!\!
\left\{ B_{lmn} + \frac{1}{\bar{n}} \left[ P(k_l)+P(k_m)+
P(k_n) \right]+ \right.
\left. \frac{1}{\bar{n}^2 } \right\} , \nn
\ea
\ba
\langle\delta_o\delta_p\delta_q\delta_r\rangle_{\rm conn.}^d  &= & (2\pi)^3 
\delta^D(\vec{k}_o+\vec{k}_p+\vec{k}_q+\vec{k}_r)\times \nn
& &\left\{ \frac{1}{\bar{n}} \left[ B_{(o+p) q r} + \mbox{perm. 
(6 terms)} \right] + \right. \nn
& &\frac{1}{\bar{n}^2} \left[ (P_{o+p+q} + \mbox{cyc. (4 terms)})+
\right.\nn
& & \left. P_{o+p}+P_{o+q}+P_{o+r} \right] +\nn
& &\left. \frac{1}{\bar{n}^3 } \right\} ,\nn
\ea
\ba
\langle\delta_1\ldots\delta_6\rangle^d_{\rm conn.} & =& 
(2\pi)^3 \delta^D(\vec{k}_1+ \ldots
+ \vec{k}_6) \times \nn
& &\left\{ \frac{1}{\bar{n}^3} \right.  [ B_{12(3+\ldots+6)}+ 
\mbox{perm. (15 terms)}+   \nn
 & &               B_{(1+2)(3+4)(5+6)}+\mbox{ perm. (15 terms.)}+ \nn
 & &               B_{1(2+3)(4+5+6)} +\mbox{perm. (60 terms)} ] + \nn
 & &  \frac{1}{\bar{n}^4 } [ P_1+\ldots+P_6 +\nn
 & &               P_{1+2} + \mbox{perm. (15 terms)}+ \nn
 & &               P_{1+2+3} + \mbox{ perm. (10 terms)}]+  \nn
 & & \left. \frac{1}{\bar{n}^5 } \right\}. \nn
\ea
The correct indices in the permutations and in the cyclical sequences 
can also be obtained using graph theory (Fry 1984, Bertschinger 1992,
Matsubara 1995).
The expression for the six-point discrete correlation function is one 
of the main original results of this work. Its expression is quite
complicated, and, it is also important to notice, the discreteness 
terms do not contribute linearly, but they are multiplied several 
times by the signal part ($B$ and $P$) and other noise parts of 
different order.

\subsection{Selection Function}

The above analysis is valid for volume-limited samples, where the
mean number density is independent of position.  In most astronomical 
catalogues, a selection criterion such as a flux limit introduces a 
position-dependent mean density, which adds complication to the 
analysis.  However, it turns out that it is very straightforward to include
a non-uniform selection function in the generating functional approach, 
by simply altering the definition of ${\cal J}$.  Although we will not 
apply these results in this paper, we present them for completeness.

Let the mean density be $\bar{n}(\vec{x})$.  We follow Feldman, 
Kaiser \& Peacock (1994; hereafter FKP) in defining a fluctuation field 
\be
F(\vx) \equiv \lambda w(\vx) [n(\vx)-\alpha n_s(\vx)] ,
\ee
where $\lambda$ is a normalisation factor, to be determined later, $n(\vx)$ is
the number density of galaxies, $n_s(\vx)$ is the number density of a 
Poisson-sampled synthetic catalogue with mean density $\bar n(\vr)/\alpha$,
and we consider the limit $\alpha\rightarrow 0$, to avoid shot noise in the
synthetic catalogue. $w(\vx)$ is an arbitrary weighting function, which may
be chosen, as in FKP for example, to minimise the variance in the 
estimate of the mean power in a shell.  

The $N$-point correlations in Fourier space are most easily
calculated by considering the process $ f=\lambda w(\vx) n(\vx)$ 
for the galaxy field and similarly for the synthetic catalogue 
$ f_s=-\alpha\lambda w(\vx) n_s(\vx)$.

The latter has to be thought as a Poisson sampling of a 
continuous and uniform underlying process whose density is 
diluted by a factor $\alpha$, therefore 
the generating functional for the process $f_s$ is:
\be
{\cal Z}_{s}({\cal J})=\exp 
\left\{ \int[\exp(-\alpha i {\cal J} \mu w) -1] 
\frac{\overline{n}}{\alpha} d^3x \right\}
\ee
and ${\cal J}$ is given by (\ref{J}) or (\ref{sourcec}).

The generating functional for the F process is then given in terms of the
generating functional for the galaxies ${\cal Z}_g$ by
\be
{\cal Z}_F({\cal J}) = {\cal Z}_{g}({\cal J})  
{\cal Z}_{s}({\cal J}) .
\label{ZofF}
\ee
The subtraction of the synthetic catalogue gives us directly 
the $N$-point correlations of the zero-mean field $F$, by the methods presented
in Section 3.2. Equation (\ref{ZofF}) includes (\ref{Z.delta}) 
as a particular case when $\alpha \rightarrow 0$.  

\subsubsection{Power spectrum and Bispectrum}

We present here the first two non-trivial terms in the expansion,
for illustration, in the case of a volume-limited sample and the general case.

Case (1): Volume-limited catalogue.  We choose 
\be 
\lambda = {1\over \bar n} ,
\ee
so that $F(\vx) = \delta(\vx)$.   In this case the transform of 
$F(\vx)$ can be used to get an unbiased estimator 
of the power spectrum:
\be
\langle |F_{\vk}|^2\rangle =
V\left[P_g(k) + {1\over \bar n}\right] \\ ,
\ee
where $V$ is the volume of the survey.
The 3-point function for a closed triangle is
\ba
\langle F_{\vk1}F_{\vk2}F_{\vk3}\rangle & =& 
 V\left[B_g(\vk_1,\vk_2,\vk_3) \right. \nn
& & \left. {1\over \bar n}(P_{g1}+P_{g2}+P_{g3})+{1\over \bar n^2}\right] .
\label{dirac}
\ea
Comparison with the infinite volume case shows that the Dirac delta function 
is replaced by $V/(2\pi)^3$.  Results for different Fourier transform 
conventions and power spectrum and bispectrum definitions are given 
in the Appendix.

Case (2): Varying selection function.  We choose
\be
\lambda^{-2} = I_{22} ,
\ee
where
\be
I_{ij} \equiv \int d^3\vx w^i(\vx) \bar n^j(\vx). 
\label{Iij}
\ee
Here we find
\be
\langle |F_{\vk}|^2\rangle = P_g(k) + {I_{21}\over I_{22}},
\ee
\ba
\!\!\!\!\!\!\!\!\!& & \langle F_{\vk1}F_{\vk2}F_{\vk3}\rangle = \nn
\!\!\!\!\!\!\!\!\!& &  {I_{33}\over I_{22}^{3/2}}
\left[B_g(\vk_1,\vk_2,\vk_3) + 
{I_{32}\over I_{33}}(P_{g1}+P_{g2}+P_{g3})+{I_{31}\over I_{33}}\right].
\ea
Comparison with the uniform cube shows that to get the $N$-point functions in
the general case the following substitutions may be made:
\be
V \rightarrow {I_{NN}\over I_{22}^{N/2}}; \qquad {1\over \bar n^q} \rightarrow
{I_{N(N-q)}\over I_{NN}}.
\ee
\section{The choice of triangles in evaluating the bispectrum}

The methods of Section 3 allow us to calculate the means and covariance
matrix for an arbitrary set of triangles, so we may use maximum 
likelihood to estimate the bias parameters (equation \ref{like}).  We are 
left with a question of which triangles to use as our data.

The choice of triangles to analyse is of course very wide.  A likelihood
analysis could include a very large number of different shapes and sizes,
provided that the full covariance matrix is calculated.  In practice,
CPU and memory considerations force one to consider a subset
of triangles, and the analysis is considerably simplified if the
covariance matrix is diagonal.  This can be achieved by ensuring that 
any $\vk$ appears in only one triangle (and, since $\delta_{-\vk}
=\delta_{\vk}^*$, one must ensure that $-\vk$ doesn't appear elsewhere
either).  This throws away some of the information,  but the number
of independent data points is at most 3 times the number one uses with
this method, so the information content may not be reduced significantly.
The choice of triangle shape is influenced by the behaviour of the
bispectrum.  As we shall see, most of the signal comes from the largest
k-vectors for which second-order perturbation theory holds.  We therefore
make the vectors as large as possible, and for this paper we choose 
only equilateral triangles.  As discussed in Section 2, a single choice
of shape does not allow us to separate $b_1$ and $b_2$.  Indeed, in the 
absence of noise, a single triangle shape determines the combination 
$b_1/(1+b_2/2Jb_1)$.    In this paper, we choose two shapes of triangle,
to lift the degeneracy.  The first is the equilateral configuration,
the second we refer to as `degenerate', and consists of a repeated vector,
and one of twice the amplitude and opposite direction.

In practice the equilateral triangles are obtained as follows:
all the $k$-vectors characterized by a polar angle less or equal to 
$60^\circ$ and a longitude angle less or equal to $180^\circ$, are considered 
to be the first vector of a triangle.
The second and the third are lying  symmetrically in a cone of 
semi-amplitude $60^\circ$ centred around $-\vec{k}_1$. Obviously 
the choice is not unique, and is made at random.
The number of actually distinct triangles that one can
obtain for a given $\vec{k}_1$ is limited by the discreteness 
of the $k$-space and increases with $|k|$.  A similar procedure is followed
for the degenerate case, but no wavevectors appear in two triangles (of either
shape).

For a discrete grid in a volume-limited cuboidal volume,  neighbouring 
triangles in $k$-space give uncorrelated estimates of the bispectrum.
This feature was checked by investigating the values taken
by the off-diagonal terms in the covariance matrix.

We have tested the method with an $N$-body simulation 
created with Couchman's (1991) $AP^3M$ code from a Cold Dark Matter (CDM)-like
simulation from the Virgo consortium (Jenkins et al. 1996) chosen to match as
closely as possible the present-day galaxy power spectrum.

In Fig. 2 we have evidence in the power spectrum for the  breakdown of
linear perturbation theory at 
$\Delta_g^2(k) \equiv k^3 P_g(k)/(2\pi^2) 
\simeq 1$, or at a wavenumber of $k \simeq 0.3$.   
Fig. 3 and 4 show the bispectrum (for the galaxy field), in the 
dimensionless form
\be
\chi ^3 \equiv \left({2\over \pi^2}\right)^2 B k_1^3 k_2^3 
\ee
and with this convention
\be
\langle \delta^3\rangle=\int d \cos \theta_{1} d \varphi _{1} 
\frac{d \ln k_1}{4 
\pi} d \cos \theta_{2} d \varphi_{2} \frac{ d \ln k_2}{4 \pi} \chi ^3 
(\vec{k}_1, \vec{k}_2). 
\ee
In these figures we see evidence of the breakdown of second-order perturbation 
at $k \sim 0.6$ for the equilateral triangle, and at a much higher $k \sim 1$ 
for the degenerate case.   The non-linear $\Delta^2(k) 
\simeq 3$ and $12$ respectively 
(cf Fry, Melott \& Shandarin 1995).  We use triangles with 
wavenumbers up to $k=0.55$ for the equilateral case, and between $0.55$ and 
$1.1$ in the degenerate case in the analysis which follows.
\begin{figure}
\begin{center}
\setlength{\unitlength}{1mm}
\begin{picture}(90,70)
\includegraphics{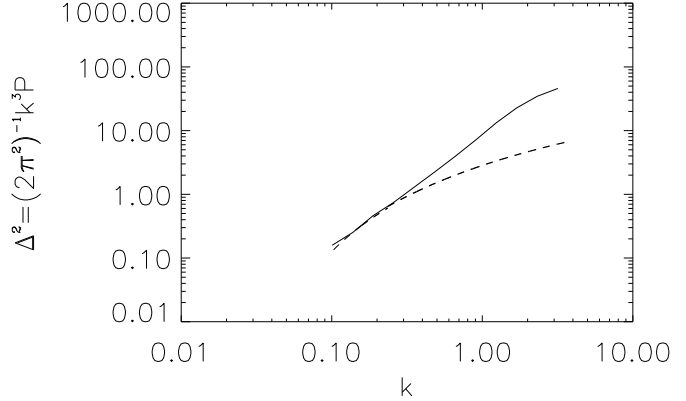}
\end{picture}
\end{center}
\caption{The power spectrum in the $N$-body simulation (solid), along with the
linear power spectrum (dashed).}
\end{figure}

\begin{figure}
\begin{center}
\setlength{\unitlength}{1mm}
\begin{picture}(90,70)
\includegraphics{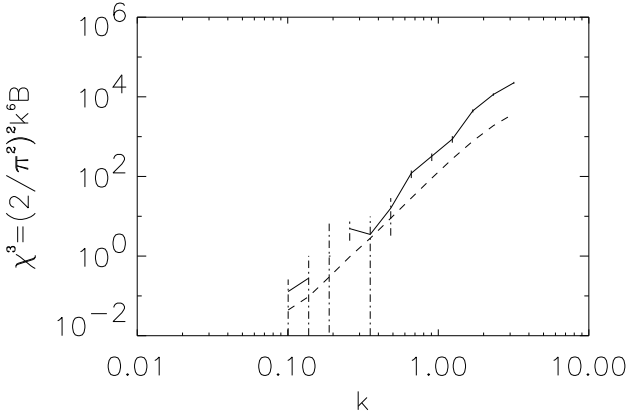}
\end{picture}
\end{center}
\caption{The measured bispectrum from the simulation (solid), with the
theoretical prediction from second-order perturbation theory shown dashed.
Equilateral triangles are used, and no wavevector appears in more than one 
triangle.   Error bars are errors in the mean for each bin.}
\end{figure}

\begin{figure}
\begin{center}
\setlength{\unitlength}{1mm}
\begin{picture}(90,70)
\includegraphics{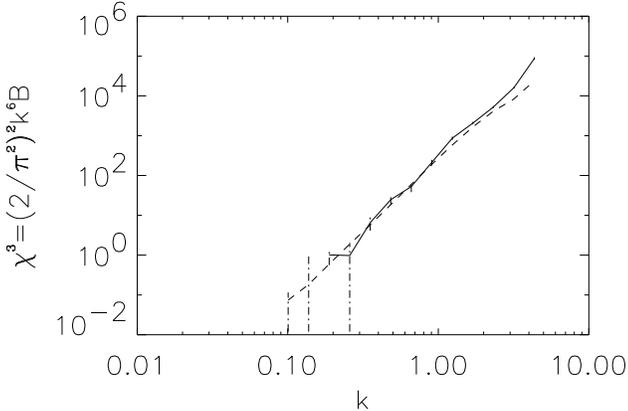}
\end{picture}
\end{center}
\caption{As in Fig. 3, but for `degenerate' triangles $k$ in the abscissa 
is the smaller wavevector.}
\end{figure}

In practice, we have used the observed power spectrum (after shot noise 
subtraction),
rather than the linear power spectrum,  to calculate the predicted 
bispectrum ($b_1=1$, $b_2=0$), shown in Fig. 3 and 4.  
This includes second-order
corrections, but for the leading-order bispectrum term, there is no
difference.   An alternative would be to linearise the power spectrum, 
as in Peacock \& Dodds (1994), but this requires knowledge of $\Omega_0$.

\subsection{Likelihood analysis}

As a first step we analyze the unbiased distribution of mass points, 
seeking to recover the values $b_1=1$ given that $b_2=0$.
    
In the case where the covariance matrix is diagonal 
the likelihood function is simplified
\begin{equation}
{\cal L}(c_1,c_2)=\frac{1}{(2 \pi )^{M/2} \Pi _{\alpha } 
\sigma_{\alpha }} \exp \left\{ {-\frac{1}{2} \sum _{\alpha } \frac
{(D_{\alpha }-\mu _{\alpha })^2}{\sigma^2 _{\alpha }}}\right\} 
\end{equation}
and very large numbers of triangles can be used.  We have used as many as
possible between the largest wavelengths in the cube and the breakdown
of second-order perturbation theory (determined above).  Fig. 5 shows
the likelihood for $b_1$, assuming that $b_2=0$, from the degenerate 
and equilateral triangles.  The correct value is
recovered, within the errors,  but the error bar in the equilateral case is 
uncomfortably large.   
Rather curiously, the error bar can be reduced by splitting the 
volume into sub-samples, as is shown in the next section.
\begin{figure}
\begin{center}
\setlength{\unitlength}{1mm}
\begin{picture}(90,70)
\includegraphics{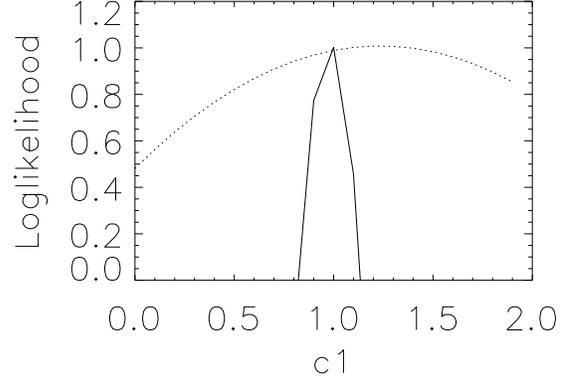}
\end{picture}
\end{center}
\caption{Likelihood of $c_1 = 1/b_1$ where $b_1$ is the linear bias parameter,
for the (unbiased) $N$-body simulation, for $0.55<k<1.1$, using degenerate 
triangles, and for equilateral triangles (dotted) for $k<0.55$.}
\end{figure}

\subsection{Expected variance in $c_1$}

We ignore shot noise for simplicity.  For equilateral triangles we have:

\begin{equation}
\mu_{\alpha}=\langle D_{\alpha}\rangle=\langle Re(\delta_{\vec{k}1} 
\delta_{\vec{k}2}
 \delta_{\vec{k}3})\rangle = c_1 \frac{12}{7}P_g^2 \delta^D \equiv c_1
\mu_\alpha^{(1)}.
\end{equation}
The variance in this approximation is
\begin{equation}
\sigma^2_{\alpha}=\langle D_{\alpha}^2\rangle-\langle D_{\alpha}\rangle^2
= {1\over 2}P_g^3 (\delta ^D)^3.
\end{equation}
The factor of 2 is the same factor as in the transition from complex $Z_\alpha$
to real $D_\alpha$ in equation (\ref{ZZstar}).  Note that of the fifteen 
signal terms
in the covariance matrix $\langle Z_\alpha Z_\beta^*\rangle$, only one 
survives.
For a discrete transform in a cube of volume $V$,
the Dirac delta function is replaced by $V/(2\pi)^3$, as in (\ref{dirac}).

Since $\sigma^2_{\alpha}$ is independent of $c_1$,  the error on $c_1$ is
\begin{equation}
\sigma_{c_1}^{-2}=-\left.\frac{\partial ^2 \ln {\cal L}}{\partial c_1^2}
\right|_{c_1=\hat{c}_1}=\sum_{\alpha}\frac{[\mu_{\alpha}^{(1)}]^2}
{\sigma_{\alpha}^2}.
\end{equation}
In practice we are dealing with more than a single mode, and moreover, 
in $k$-space the modes form a discrete set, with density of states
$g={V/(2 \pi )^3}$.  The number of uncorrelated triangles 
in a thin shell of width 
$\delta (\ln k)$ is:
\begin{equation}
\frac{1}{2} 4 \pi k^3 \delta (\ln k) g \frac{1}{3} ,
\end{equation}
where the factor $\frac{1}{2}$ arises from the reality of the 
density fluctuation field and the factor $\frac{1}{3}$ from the 
requirement of uncorrelated data (no wavevectors in common between 
any two triangles). 

Considering contributions from all the shells,
and moving to the continuous limit this becomes:
\begin{equation}
\sigma^{-2}_{c_1}= \frac{V}{12 \pi ^2}\int _{k_{\rm min}}^{k_{\rm max}}  
d(\ln k)
\frac{\left[\mu(\vec{k})^{(1)}\right]^2}{\sigma^2({k})} k^3,
\end{equation}
where $k_{\rm max}$ is set by the breakdown of perturbation theory,
and $k_{\rm min}$ is set by the size of the sample.

Inserting the appropriate expressions 
in the previous equation one obtains:
\begin{equation}
\sigma^{-2}_{c_1}={48\over 49}\sigma_0^2(k_{\rm max}),
\end{equation}
where
\begin{equation}
\sigma_0^2(k_{\rm max})= {1\over 2\pi^2} \int ^{k_{\rm max}}_{k_{\rm min}} 
d(\ln k) P_g k^3 
\end{equation}
is the variance in the galaxy field for a top hat filter in $k$-space 
truncated at  $k_{\rm max}$.  It is also truncated at the sample size at
low-$k$, which makes the definition slightly non-standard.  Hence
\begin{equation}
\sigma _{c_1} \simeq {1.01\over {\sigma_0(k_{\rm max})}} ,
\end{equation}
which gives the error of 1.48 seen in Fig. 5, for the
cutoff at $k=0.55$. 
A curious feature of this analysis is that the error is
independent of the size of the survey, motivating the splitting of
the volume into independent smaller units.  This loses some long-wavelength
modes, but these contain little information anyway.  Fig. 6 shows the
likelihood curves for several sub-volumes,  and comparison with Fig. 5 
demonstrates that the
error is independent of the size of the subvolume.  This is somewhat 
reminiscent of the equally counter-intuitive result that the variance in 
an estimate of the power spectrum is also independent of the size of the
survey, motivating a splitting of the volume as a 
possible way to increase signal-to-noise in that case (Press et al. 1992).  

\begin{figure}
\begin{center}
\setlength{\unitlength}{1mm}
\begin{picture}(90,70)
\includegraphics{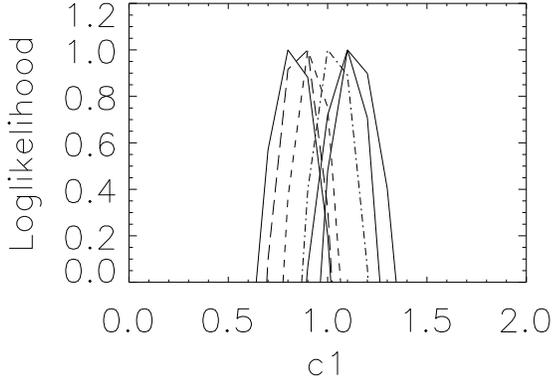}
\end{picture}
\end{center}
\caption{Likelihood of $c_1 = 1/b_1$
for several sub-volumes, each containing 1/8 the volume, for degenerate 
triangles for the same $N$-body 
simulation as Fig. 4.}
\end{figure}

\subsection{Determining $b_1$ and $b_2$}

For simplicity, the error analysis of the previous section assumed that
the quadratic bias term $b_2$ is zero.  In practice,  this may not be
true, and we need to estimate both $b_1$ and $b_2$ from the data.
To lift the degeneracy between $b_1$ and $b_2$, we need to consider
at least two triangle configurations.   The optimal configurations
are not easy to determine, and will certainly depend on the power
spectrum, as anything other than the equilateral configuration 
involves wavenumbers of different magnitude.  In addition, each
triangle shape leads to approximate determination of a degenerate
combination of $b_1$ and $b_2$, so ideally we would like these 
degenerate lines to cross at large angle.   Adding the degenerate triangles 
to the equilateral configuration is particularly effective, as each 
degenerate triangle
removes only two wavevectors from the available set, and also second-order
perturbation theory appears to work well into the nonlinear regime.
Fig. 7 shows the joint likelihood for $c_1$ and $c_2$ for the limits on
$k$ previously stated.  The true solution is nicely within the error
bars.

\begin{figure}
\begin{center}
\setlength{\unitlength}{1mm}
\begin{picture}(90,70)
\includegraphics{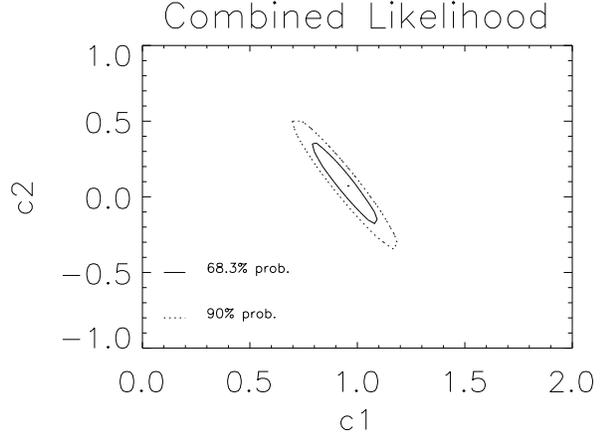}
\end{picture}
\end{center}
\caption{Joint likelihood of $c_1 = 1/b_1$ and $c_2 \equiv b_2/b_1^2$.  
Contours contain 68.3 and 90 percent of the {\em a posteriori} probability 
assuming the likelihood is a bivariate gaussian, and the prior 
probabilities for $c_1$ and $c_2$ are uniform.
}
\end{figure}

\section{Discussion}

We have presented here the first steps towards the goal of using the
bispectrum to measure the linear bias parameter $b$ from large-scale structure 
data.   The aim of this is ultimately to combine measurements of $b$
with estimates of $\beta = \Omega_0^{0.6}/b$ from redshift distortion
studies to get an unambiguous estimate of the density parameter $\Omega_0$.
In this paper we have presented a method for calculating the covariance
matrix for bispectrum estimates, thus allowing maximum likelihood to be
used to estimate $b$, and tested the methods on $N$-body simulations.   
The noise calculation is a general method which will be useful for 
calculating the properties of any high-order statistics, and is the 
major new advance of this paper.  We have also shown how the 
signal-to-noise of the bias estimate may 
be improved by splitting the volume into subsamples.   

It is worth making some remarks about the shape information,  as
this is crucial in lifting a degeneracy which exists physically between
the effect of gravitational evolution and nonlinear bias (in the analysis,
this manifests itself as a degeneracy between the linear and quadratic 
bias terms).   In essence, we are exploiting the fact that gravitational
instability skews the density field away from its initial Gaussian symmetry.
However, this can also be achieved by a nonlinear bias, so we need a 
further property to distinguish the two.     It is the shape information which
allows this.  Although we work in Fourier space, the argument is most
easily stated in real space.  The effect of a nonlinear bias, operating
on the Gaussian field, is to 
shift iso-density contours up (or down),  but maintaining the shape of the
contour.  Gravitational evolution, however, will change the shapes,
usually leading to flattening of collapsing structures (e.g. 
Lin, Mestel \& Shu 1965).   By utilising shape information, we can, in 
principle, decide to what extent nonlinear biasing and gravitational
evolution are responsible for the skewed density field.

A number of unresolved issues remain, and these will be addressed in 
a subsequent paper.  There are two major ones:   the practical effects of
a varying selection function, and the effect of redshift-space distortions,
arising from the fact that the redshift is an imperfect 
distance indicator because of the contaminating effects of peculiar
velocities.  This effect can be regarded as being split into two components:
a large-scale distortion resulting from coherent inflow into overdense 
regions, and the `Fingers-of-God' arising from virialised, highly 
non-linear structures.   Since we have argued that it is effective to use
subsamples, we would expect to be able to use Kaiser's (1987) distant-observer 
approximation, for a deep survey, to model the large-scale distortions
(see also Hivon et al. 1995).
Of more concern is that most of the signal used to measure the bias parameters
comes from the mildly non-linear regime, and it may be that simple
models for the Fingers-of-God (such as the incoherent velocity dispersion
introduced by Peacock (1992)) may not be adequate for our purposes.  

Assuming these problems can be dealt with, the accuracy of the 
determination of $b$ will be dependent on the number of
subvolumes we make, and this is determined by how small the subvolumes 
may be.   From the numerical simulations, we find that the subvolumes
degrade significantly the bispectrum estimate if the size of the box
is less than about the nonlinear wavelength $2\pi/k_{\rm NL}$, or about
20 $h^{-1}$ Mpc for IRAS galaxy data.  One would therefore expect rather 
weak constraints on $b$ from the IRAS PSCz survey, but a much improved error
of under 10\% from the Sloan digital sky survey or Anglo-Australian 2 
degree-field survey.  These are 
rough estimates, based on the shot noise and depth of each survey; 
it will require more detailed study to see if this accuracy can be 
achieved or surpassed.

\section* {Acknowledgments}

LV acknowledges the support of PPARC and the Dewar \& Ritchie fund of the
University of Edinburgh.  Computations were made using STARLINK facilities.
AFH and LV thank the Department of Astronomy in Padova for hospitality.  
SM thanks the University of Edinburgh for hospitality.   We thank 
John Peacock, Sergei Shandarin, Francesco Lucchin, Lauro Moscardini, 
Roman Scoccimarro  and
Paolo Catelan for useful discussions.

\appendix

\section{Fourier transform conventions}

There are many possibilities for placing of factors of $2\pi$ etc. in
the definitions of the power spectrum, bispectrum and Fourier transform.
These are characterised by the constants $A,D$ and $E$ below.  Our choice 
is $A=1$, $D=E=(2\pi)^3$, which seems to yield the minimum complexity 
in the formul\ae.  With the addition of $E$, this is the same convention as 
FKP, but differs from Bertschinger (1992), who uses $A=(2\pi)^{-3}$, $D=E=1$.
\ba
\delta_{\vk} & \equiv & A\int d^3x \delta(\vx) \exp(-i\vk.\vx)\nn
\langle \delta_{\vk 1} \delta_{\vk 2} \rangle & \equiv & D P(k) \delta^D 
(\vk_1+\vk_2)\nn
\langle \delta_{\vk 1}\delta_{\vk_2}\delta_{\vk_3} \rangle & \equiv & 
E B(\vk_1,\vk_2,\vk_3) \delta^D(\vk_1+\vk_2+\vk_3).\nn
\ea
We present below some of the simpler results, including shot noise, 
with general Fourier conventions for workers whose convention is 
different from ours.
For the case of a varying selection function, the power is
\be
\langle |F_{\vk}|^2\rangle = \lambda^2\left[{P(k) D I_{22}\over (2\pi)^3}+ 
A^2 I_{21}\right]
\ee
and the 3-point function is 
\ba
\langle F_{\vk1}F_{\vk2}F_{\vk3}\rangle & = & \lambda^3
\left[{E I_{33}\over (2\pi)^3} B({\vk}_1,{\vk}_2,{\vk}_3) + \right.\nn
& & \left. 
{A D I_{32}\over (2\pi)^3}(P_1+P_2+P_3)+A^3 I_{31}\right].\nn
\ea
The conventions may be related to the inverses
\ba
P(k) & = & {(2\pi)^3 A^2\over D} \int d^3x \xi^{(2)}(x) \exp(-i\vk.\vx)\nn
B({\vk}_1,{\vk}_2,{\vk}_3) & = & {(2\pi)^3 A^3\over E} \int d^3x_{12} d^3x_{23}
\nn
& & \!\!\!\!\!\!\!\!\!\!\!\!\!\!\!\!\!\!\!\!\!
\xi^{(3)}(\vx,\vx+\vx_{12},\vx+\vx_{23}) 
\exp(-i\vk_2.\vx_{12}-i\vk_3.\vx_{13})\nn
\ea
(the latter for a closed triangle).
The source function for the generating functional in Fourier space is modified
in the continuous case from (\ref{sourcec}) to:
\be
{\cal J}_{\vec{k}}(\vec{x})=A\sum_{m=1}^N s_m e^{-i \vec{k} \cdot \vec{x}}
\widetilde{W}(\vec{k}_m) .
\ee

\end{document}